\documentclass[12pt,preprint]{aastex}
\usepackage[]{natbib}
\usepackage[american]{babel}
\usepackage{epsfig}
\usepackage{amsmath}
\usepackage{amssymb}
\usepackage{subfigure}
\usepackage{graphicx}

\begin{document}
\title{A search for a dark matter annihilation signal\\ towards the Canis Major overdensity with H.E.S.S.}

\author{F.~Aharonian\altaffilmark{1,13}, A.G.~Akhperjanian\altaffilmark{2}, U.~Barres de Almeida\altaffilmark{8,\ast}, A.R.~Bazer-Bachi\altaffilmark{3}, B.~Behera\altaffilmark{14}, W.~Benbow\altaffilmark{1}, K.~Bernl\"ohr\altaffilmark{1,5}, C.~Boisson\altaffilmark{6}, V.~Bochow\altaffilmark{1}, V.~Borrel\altaffilmark{3}
% ...
}
\altaffiltext{1}{Max-Planck-Institut f\"ur Kernphysik, P.O. Box 103980, D 69029 Heidelberg, Germany} 
\altaffiltext{2}{Yerevan Physics Institute, 2 Alikhanian Brothers St., 375036 Yerevan,Armenia} 
\altaffiltext{3}{Centre d'Etude Spatiale des Rayonnements, CNRS/UPS, 9 av. du Colonel Roche, BP 4346, F-31029 Toulouse Cedex 4, France} 
% ...
\altaffiltext{5}{Institut f\"ur Physik, Humboldt-Universit\"at zu Berlin, Newtonstr. 15, D 12489 Berlin, Germany} 
\altaffiltext{6}{LUTH, Observatoire de Paris, CNRS, Universit\'e Paris Diderot, 5 Place Jules Janssen, 92190 Meudon, France} 
% ...
\altaffiltext{8}{University of Durham, Department of Physics, South Road, Durham DH1 3LE, U.K.} 
% ...
\altaffiltext{13}{Dublin Institute for Advanced Studies, 5 Merrion Square, Dublin 2, Ireland} 
\altaffiltext{14}{Landessternwarte, Universit\"at Heidelberg, K\"onigstuhl, D 69117 Heidelberg, Germany}
% ...
\altaffiltext{$\ast$}{supported by CAPES Foundation, Ministry of Education of Brazil}
% ...

\begin{abstract}
A search for a dark matter (DM) annihilation signal into $\gamma$-rays toward the direction of the Canis Major (CMa) overdensity is presented. The nature of CMa is still controversial and one scenario represents it as a dwarf galaxy, making it an interesting candidate for DM annihilation searches. A total of 9.6 hours of high quality data were collected with the H.E.S.S. array of Imaging Atmospheric Cherenkov Telescopes (IACTs) and no evidence for a very high energy $\gamma$-ray signal is found. Upper limits on the CMa dwarf galaxy mass of the order of 10$^{9}$ M$_{\odot}$ are derived at the 95$\%$ C.L. assuming neutralino masses in the range 500 GeV - 10 TeV and relatively large annihilation cross-sections. Constraints on the velocity-weighted annihilation cross section $\langle\sigma v\rangle$, are calculated for specific WIMP scenarios, using a NFW model for the DM halo profile and taking advantage of numerical simulations of hierarchical structure formation. 95$\%$ C.L. exclusion limits of the order of 5 $\times$ 10$^{-24}$ cm$^{3}$ s$^{-1}$ are reached in the 500 GeV - 10 TeV DM particle mass interval, assuming a total halo mass of 3 $\times$ 10$^{8}$ M$_{\odot}$.
\end{abstract} 


\section{Introduction}
It is widely believed that around one third of the energy content of the Universe is made of cold matter, of which only a small fraction is luminous and is of baryonic origin. In the framework of the Cold Dark Matter (CDM) scenarios, most of the matter is composed of non-baryonic Weakly Interacting Massive Particles (WIMPs). The standard model of particle physics does not provide a natural and suitable candidate for the Dark Matter (DM) particle and many theories beyond the standard model have been proposed to explain its origin and properties (see \citep{Bertone} for a recent review). In various models, the self-annihilation of WIMPs gives a $\gamma$-ray continuum emission resulting from the hadronization of primary annihilation products. The spectral features of such a DM annihilation radiation might help to distinguish it from ordinary astrophysical sources \citep{Bergstrom1,Bergstrom2}. The indirect detection of DM through very high energy (VHE) $\gamma$-rays is one of the best ways to probe the astrophysical nature of the DM and the H.E.S.S. array of Cherenkov telescopes, dedicated for detection of VHE $\gamma$-rays in the 100 GeV-10 TeV energy regime, is an interesting instrument for this purpose.\\
Many astrophysical objects, ranging from DM clumps to galaxy clusters are expected to lead to DM particle annihilation signals detectable with sufficiently sensitive instruments. Regions of high concentration of DM are good candidates to search for such annihilations and the Galactic Centre (GC) was first considered. H.E.S.S. observations of the GC region \citep{SgrA*} revealed a source of VHE $\gamma$-ray emission (HESS J1745-290) but ruled out the bulk of the signal as of DM origin \citep{Rolland}. Prospects for indirect detection in the elliptical galaxy M87 at the center of the Virgo cluster were also investigated. The time variability of the VHE $\gamma$-ray signal observed by H.E.S.S. \citep{M87} gave clear evidence that the signal was not of a sole DM origin. There are also other candidates with high DM density in relative proximity that might lead to detectable DM annihilation signals. Satellite dwarf galaxies of the Milky Way (MW) such as Sagittarius, Draco or Canis Major are popular targets, owing to their relatively low astrophysical background \citep{EvansFerrer}. Indeed, dwarf spheroidal galaxies usually consist of stellar populations with no hot or warm gas and no cosmic rays, and are among the most extreme DM-dominated environments. A null result concerning the search for DM toward the Sagittarius dwarf spheroidal galaxy (Sgr dSph) direction was published by the H.E.S.S. collaboration \citep{Moulin}. Constraints on the parameter spaces of two popular WIMPs models, namely the R-parity conserving Minimal Supersymmetric extension of the Standard Model (MSSM) and Kaluza-Klein (KK) scenarios with K-parity conservation, were derived. A null result was also established by the MAGIC collaboration \citep{MAGIC} and the WHIPPLE collaboration \citep{WHIPPLE}, when searching for a DM annihilation signal toward the Draco dwarf galaxy. Upper limits on the velocity-weighted annihilation cross-section of DM particle were derived in the framework of minimal SUper GRAvity (mSUGRA) models and MSSM models, respectively.\\
The present paper reports the search for a DM annihilation signal towards the direction of the CMa overdensity with the H.E.S.S. array of Cherenkov telescopes. The paper is organized as follows: in Sec 2 the controversial nature of the CMa overdensity is briefly discussed; in Sec 3 the analysis of the data is presented, while in Sec 4 the predictions for DM annihilation into $\gamma$-rays in the CMa overdensity are discussed. Constraints on the WIMP velocity-weighted annihilation rate, as well as on the CMa total mass, are given.  

\section{A Galactic warp or the relic of a dwarf galaxy?}
Since its discovery \citep{Martin}, the nature of the Canis Major (CMa) overdensity is the subject of many discussions over whether it is a dwarf galaxy or simply a part of the warped Galactic disk. According to \citep{Momany}, the CMa overdensity simply reflects the warp and flare of the outer disk, a structure frequently observed in spiral galaxies such as the MW. The comparison of the kinematics of the CMa stars with those of the Galactic thick disk shows that the CMa stars do not have peculiar proper motions, which cast some doubts on the dwarf Spheroidal (dSph) nature of the CMa overdensity. The second scenario, which is of interest for the aim of this paper, considers this elliptical overdensity as the remnant of a disrupted dwarf galaxy that could have created the Monoceros ``ring'' structure \citep{Martin}. Indeed, numerical simulations show that such a structure can be explained by an in-plane accretion event, in which the remnant of the dwarf galaxy would have an orbital plane close to the Galactic plane. Another argument in favor of this scenario is that the CMa star populations do not exhibit the same properties as those in the disk since they have a relatively low metallicity \citep{deJong}. In this case, the CMa star population would not belong to the Galactic disk, in contrast to what was found in \citep{Momany}. The mass, luminosity and characteristic dimensions of CMa appear quite similar to those of the Sgr dwarf galaxy. As for many dSph, the CMa overdensity would thus be an interesting candidate for DM detection. In the remainder of the paper, the Canis Major object is assumed to be a dwarf galaxy.\\
The CMa overdensity is located towards the Galactic anti-centre direction at roughly 8 kpc from the sun \citep{Bellazinni} and is the closest observed dwarf galaxy. It is a very extended object ($\mathrm{\Delta}$l = 12$^{\mathrm{\circ}}$, $\mathrm{\Delta}$b = 10$^{\mathrm{\circ}}$) with a roundish core approximately centered at l = 240$^{\mathrm{\circ}}$ and b = -8$^{\mathrm{\circ}}$ according to various star surveys in this region \citep{Martin, Martinez}. In contrast to the Sgr dSph, neither dispersion velocity measurements, nor luminosity profiles are available so that an accurate modelling of the CMa DM halo profile is not possible. However, there are enough constraints to estimate the expected $\gamma$-ray flux from DM particle annihilations in this object. The annihilation cross-section is given by the particle physics model (see section IV). As concerns the mass content of CMa, the narrow dispersion between the average mass values found for different dSph galaxies in the local group \citep{Mateo, Walker} is an indication that dSph's may possibly have a universal host halo mass \citep{Dekel}. The mass of the CMa dwarf galaxy can then be inferred to be in the same range as the Sgr dwarf galaxy and many other dSph's so that the CMa total mass would range between 10$^{8}$ and 10$^{9}$ M$_{\odot}$ \citep{Martin}. For instance, reference \citep{EvansFerrer} gives a model where the CMa mass is taken as 3 $\times$ 10$^{8}$ M$_{\odot}$. The H.E.S.S. large FoV covers a large part of the CMa core, optimizing the chances to see a potential DM annihilation signal. 

\section{H.E.S.S. observations and analysis}

\subsection{The H.E.S.S. array of Imaging Atmospheric Cherenkov Telescopes}
H.E.S.S. is an array of four Imaging Atmospheric Cherenkov Telecopes (IACT's) \citep{Hofmann} located in the Khomas Highland of Namibia at an altitude of 1800 m above sea level. The instrument uses the atmosphere as a calorimeter and images electromagnetic showers induced by TeV $\gamma$-rays. Each telescope collects the Cherenkov light radiated by particle cascades in the air showers using a large mirror area of 107 m$^2$ and a camera of 960 photomultiplier tubes (PMT's).  The four telescopes are placed in a square formation with a side length of 120 m. This configuration allows for a accurate reconstruction of the direction and energy of the $\gamma$-rays using the stereoscopic technique. The cameras cover a total field of view of 5$^\circ$ in diameter. The energy threshold of the H.E.S.S. instrument is approximately 100 GeV at zenith and its sensitivity allows to detect fluxes larger than 2 $\times$ 10$^{-13}$ cm$^{-2}$ s$^{-1}$ above 1 TeV in 25 hours. More details on the H.E.S.S. experiment can be found in \citep{HESS}.

\subsection{Data processing}
\begin{figure}
\plottwo{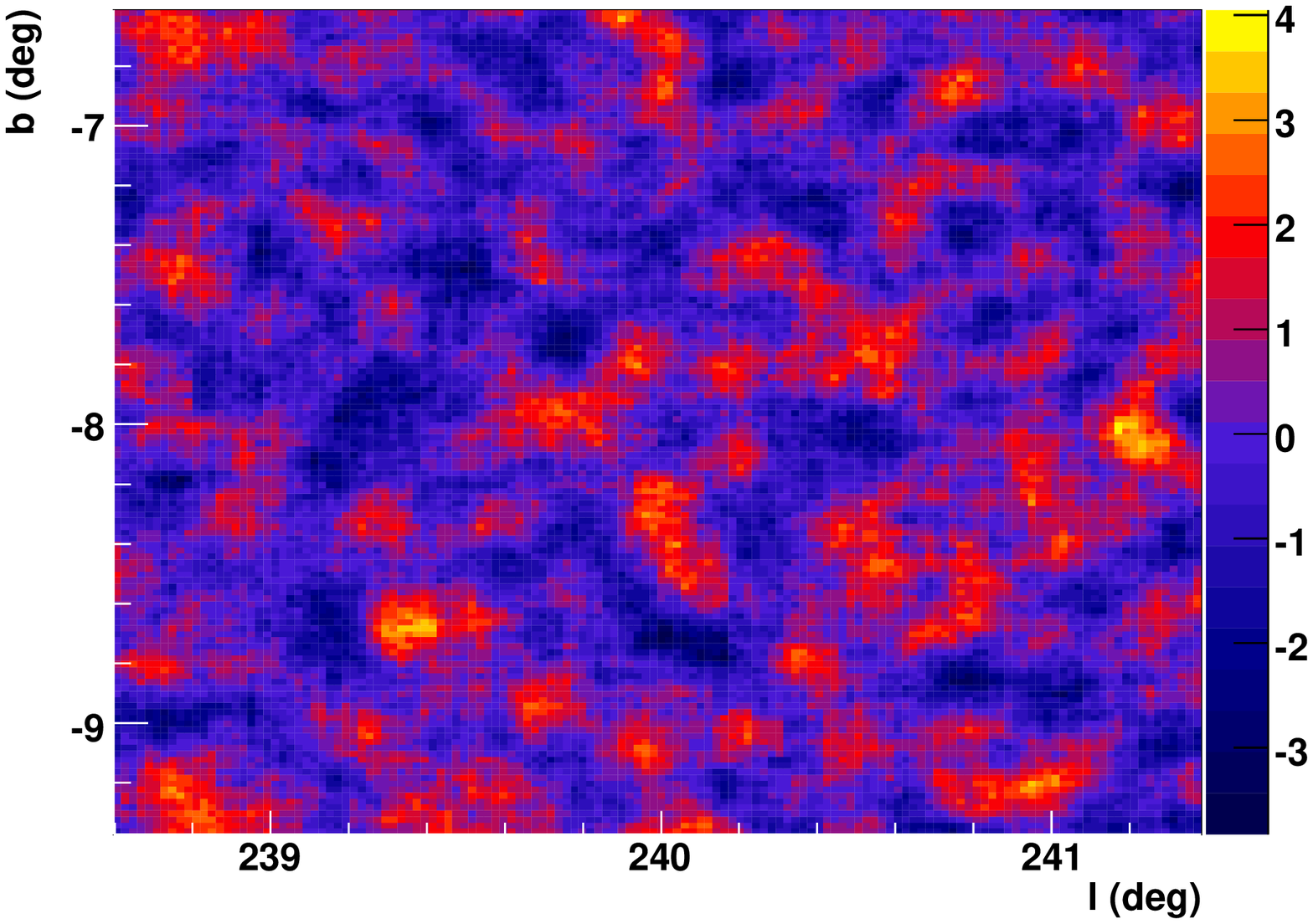}{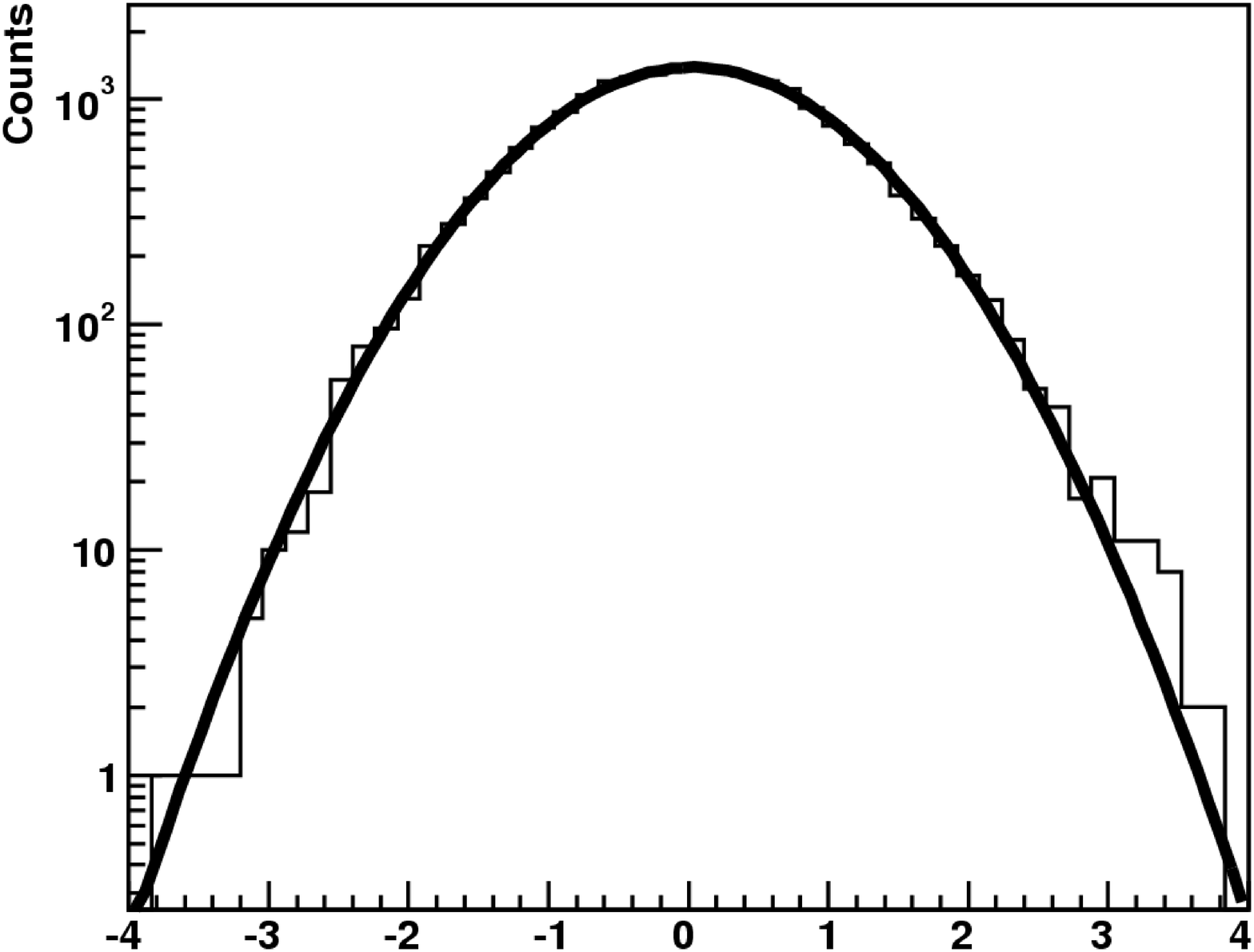}
\caption{(left) Significance map corresponding to the excess map computed in the analysis (see text), calculated according to the Li $\&$ Ma method \citep{LiMa}. (right) Significance distribution derived from the significance map. The solid line shows the Gaussian fit. The mean value is 0.01 $\pm$ 0.004 and the corresponding variance is 1.000 $\pm$ 0.005.}\label{fig1}
\end{figure}

Observations of the CMa dwarf galaxy with H.E.S.S. were carried out in November 2006 with pointing angles close to the zenith and extending up to 20$^\circ$. The nominal pointing direction was l = 240.15$^{\mathrm{\circ}}$ and b = -8.07$^{\mathrm{\circ}}$ in Galactic coordinates. The data were taken in ``wobble mode'' with the telescope pointing typically shifted by $\pm$0.7$^{\mathrm{\circ}}$ from the nominal target position \citep{Berge}. The wobble observing mode reduces the systematic effects in the background estimates. The dataset used for image analysis was selected using the standard quality criteria, excluding runs taken under bad or variable weather conditions. The CMa dataset amounts to 9.6 hours of live time after quality selection.\\
Two different techniques are combined for the data processing. The first technique computes the ``Hillas geometrical moments'' of the shower images to reconstruct shower geometry and energy, and to discriminate between $\gamma$-ray and hadronic events \citep{Hillas}. The second technique uses a semi-analytical model of air showers which predicts the expected intensity in each camera pixel \citep{Model}. Here, the shower direction, the impact point and the primary particle energy are derived with a likelihood fit of the shower model to match the images. Standard cuts for $\gamma$-ray/hadron separations are derived by simulations. Both analyses provide an energy resolution of 15$\%$ and an angular resolution better than 0.1$^\circ$. The combination of these two techniques, referred hereafter as ``Combined Hillas/Model analysis'', uses a combined estimator (the so-called ``Combined cut'') and provides an improved background rejection. The background is estimated following the template background method. This method uses events that fail the $\gamma$-ray selection cuts. The template background modelling is well suited for the detection of sources positioned anywhere in the FoV, as it estimates the background in each sky direction. More details on this background subtraction method are given in \citep{Template}. A charge cut of 60 photoelectrons on the image size is applied, as well as a cut on the primary interaction depth of particles to optimize the signal-to-noise ratio. Events that pass the analysis cuts are labelled as ``$\gamma$ candidates'' and are stored in the so-called $\gamma$ candidate map $n_{\mathrm{\gamma}}^{\mathrm{candidate}}(l,b)$. Events that do not pass the analysis cuts are defined as ``background events'' and are stored in the so-called background map $n_{\mathrm{bck}}(l,b)$.  Table \ref{table0} shows the different cut values used to select the $\gamma$-ray events.\\
\placetable{table0}
The 2.5$^{\circ}$$\times$2.5$^{\circ}$ excess sky map is obtained by the following equation:
\begin{equation}
 n_{\mathrm{\gamma}}^{\mathrm{excess}}(l,b) = n_{\mathrm{\gamma}}^{\mathrm{candidate}}(l,b) - \alpha(l,b) \times n_{\mathrm{bck}}(l,b)\label{EqTemplate1},
\end{equation}
where $\alpha(l,b)$ refers to the template normalisation factor as described in \citep{Template}.
To search for a gamma-ray signal, the raw fine-binned maps are integrated with a 0.1$^{\circ}$ radius around each point to match the H.E.S.S. angular resolution, resulting in new oversampled maps of gamma-ray candidates and background events, and a corresponding gamma-ray excess map. Using the prescription of Li and Ma \citep{LiMa} to derive the significance for each point of the oversampled map on the basis of the gamma-ray candidate and background counts and the template normalization factor, no significant excess is found at the target position or at other points in the field of view (Fig. \ref{fig1} left panel). The distribution of significances for the entire map is shown in the right panel of Fig. \ref{fig1} and is fully consistent with statistical fluctuations of the background signal. As the excess map does not show any signal, an upper limit on the number of gamma-ray events for each point in the map can be derived using the method of Feldman and Cousins \citep{FeldmanCousins}. The uncorrelated $\gamma$ candidate and normalized background maps, plotted on a 0.2$^{\circ}$$\times$0.2$^{\circ}$ grid to have bins not smaller than the H.E.S.S. angular resolution, are used for the upper limits calculations.
\placefigure{fig1}
\section{Predictions for Dark Matter annihilations in the Canis Major overdensity}
The DM particles are expected to annihiliate into a continuum of $\gamma$-rays through various processes such as the hadronization of quark final states, hadronic decay of $\tau$ leptons and subsequent decay of mesons. Two DM candidates are commonly discussed in literature: the so-called neutralino arising in supersymmetric extensions of the standard model (SUSY) \citep{SUSY}, and the first excitation of the hypercharge gauge boson in Universal Extra Dimension theories (UED) called the B$^{(1)}$ particle \citep{KK}. Typical masses for these DM candidates range from 50 GeV to several TeV. The value of the annihilation cross-section is constrained to give a thermal relic abundance of WIMPs that is in agreement with the WMAP+SDSS derived value \citep{WMAPSDSS}. The velocity-weighted annihilation cross-sections can be as low as 10$^{-30}$ cm$^{3}$ s$^{-1}$, for scenarios involving co-annihilations processes, and be as high as 10$^{-25}$ cm$^{3}$ s$^{-1}$.\\
The expected flux $\phi_{\mathrm{\gamma}}$ of $\gamma$-rays from WIMP annihilations occuring in a spherical dark halo is commonly written as a product of a particle physics term (d$\Phi^{\mathrm{PP}}/\mathrm{d}E_{\mathrm{\gamma}}$) and an astrophysics term ($f^{\mathrm{AP}}$):
\begin{equation}
  \phi_{\mathrm{\gamma}} = \frac{\mathrm{d}\Phi^{\mathrm{PP}}}{\mathrm{d}E_{\mathrm{\gamma}}} \times f^{\mathrm{AP}}\label{Eq1}
\end{equation}
The expected number of $\gamma$-ray is then given by:
\begin{equation}
  N_{\mathrm{\gamma}} = T_{\mathrm{ON}} \times \int^{\mathrm{m}_{\mathrm{DM}}c^{\mathrm{2}}}_{\mathrm{0}} \overline{A_{\mathrm{eff}}}(E_{\mathrm{\gamma}})\phi_{\mathrm{\gamma}}(E_{\mathrm{\gamma}}) \mathrm{d}E_{\mathrm{\gamma}},\label{Eq2}
\end{equation}
where $T_{\mathrm{ON}}$ denotes the ON-source exposure time, which depends on the pointing direction, and $\overline{A_{\mathrm{eff}}}(E_{\mathrm{\gamma}})$ the averaged H.E.S.S. acceptance during data collection. The velocity-weighted cross-section for WIMP annihilation $\langle\sigma v\rangle$ and the WIMP mass are fixed to compute the particle physics term in Eq.\ref{Eq1}:
 \begin{equation}
  \frac{\mathrm{d}\Phi^{\mathrm{PP}}}{\mathrm{d}E_{\mathrm{\gamma}}} = \frac{\langle\sigma v\rangle}{4\pi m_{\mathrm{DM}}^{\mathrm{2}}} \big\lgroup\frac{\mathrm{d}N}{\mathrm{d}E_{\mathrm{\gamma}}}\big\rgroup_{\mathrm{DM}},\label{Eq3}
\end{equation}
where $(\mathrm{d}N/\mathrm{d}E_{\mathrm{\gamma}})_{\mathrm{DM}}$ is the $\gamma$-ray spectrum originating for DM particle annihilation. The shape of the continuum $\gamma$-ray spectrum predicted in the framework of the phenomenological Minimal Supersymmetric extension of the Standard Model (pMSSM) depends on the model in a complicated way. A simplified parametrization of this shape, for higgsino-like neutralinos mainly annihilating via pairs of W and Z gauge bosons, was taken from \citep{Bergstrom3}. In the case of KK B$^{(1)}$ particle annihilations, the branching ratios to final states are independent of the WIMP mass. The differential photon continuum has been simulated with the PYTHIA package \citep{PYTHIA} using branching ratios from \citep{KK}.\\
The astrophysics term $f^{\mathrm{AP}}$ illustrates the DM concentration dependency of the expected $\gamma$-ray flux toward the pointed source:
\begin{equation}
  f^{\mathrm{AP}} = \int_{\mathrm{\Delta\Omega}}\int_{\mathrm{los}}\rho^{\mathrm{2}}(l)\mathrm{d}l \mathrm{d}\Omega,\label{Eq4}
\end{equation}
where $\rho(l)$ is the mass density profile of the CMa dwarf galaxy and $\mathrm{\Delta\Omega}$ the detection solid angle ($\mathrm{\Delta\Omega}$ = 10$^{-5}$ sr, corresponding to the integration radius of 0.1$^{\circ}$).
\subsection{Model of the Canis Major Dark Matter halo within the $\Lambda$CDM cosmology}
The purpose of this section is to explain how the astrophysical term $f^{\mathrm{AP}}$ (Eq. \ref{Eq1}, Eq. \ref{Eq4}) was calculated as a function of the total CMa dark halo mass. The estimate of the astrophysical term $f^{\mathrm{AP}}$ relies on the modelling of the CMa DM mass distribution. Observationally, the DM mass content of dSph galaxies can be derived using velocity dispersion measurements of their stellar population as well as their luminosity profile. The comparison between models and observations can constrain the parameters of their assumed density profiles. In the case of the CMa dSph, the lack of available observational data prevents the modelling of its density profile in the same way as in the literature \citep{EvansFerrer, Colafrancesco, Moulin}.\\
In the absence of observational data, a standard cusped NFW halo \citep{NFW} was assumed to model the CMa dwarf mass distribution:
\begin{equation}
\rho_{\mathrm{cusped}}(r) = \frac{\rho_\mathrm{0}}{\frac{r}{r_\mathrm{s}}(1+\frac{r}{r_\mathrm{s}})^{\mathrm{2}}}, 
\end{equation}
where $\rho_\mathrm{0}$ is the overall normalisation and $r_\mathrm{s}$ the scale radius. The parameters  $\rho_\mathrm{0}$ and $r_\mathrm{s}$ determining the shape of the profile as well as the halo virial mass $M_{\mathrm{vir}}$ are found by solving the following system of 3 equations:
\begin{eqnarray}
&&M_{\mathrm{vir}} = \int_{\mathrm{0}}^{R_{\mathrm{vir}}} \rho_{\mathrm{cusped}}(r) \mathrm{d}^{\mathrm{3}}\vec{r}\label{eq1}\\
&&M_{\mathrm{vir}} = \frac{4\pi}{3}\rho_{\mathrm{200}} \times R_{\mathrm{vir}}^{\mathrm{3}}\label{eq2}\\
&&C_{\mathrm{vir}}(M_{\mathrm{vir}},z) = \frac{c_\mathrm{0}}{1+z}\times \big\lgroup\frac{M_{\mathrm{vir}}}{10^{\mathrm{14}}h^{\mathrm{-1}}M_{\odot}}\big\rgroup^{\mathrm{\alpha}}\label{eq3},
\end{eqnarray}
where $R_{\mathrm{vir}}$ is the halo virial radius. The virial radius is computed given the virial mass $M_{\mathrm{vir}}$ and is defined as the radius within which the mean density equals $\rho_{\mathrm{200}}$ ($\mathrm{\rho_{\mathrm{200}} = 200 \times \rho_{\mathrm{c}}}$, where $\rho_{\mathrm{c}}$ is the critical density of the universe \footnote{\url{http://pdg.lbl.gov}}).\\ 
Eq. \ref{eq3} relates the concentration parameter $C_{\mathrm{vir}}$ to the virial mass. The concentration parameter is defined as the ratio between the virial radius and the scale radius $R_{\mathrm{vir}}/r_\mathrm{s}$ in the case of a NFW profile. The relation between $C_{\mathrm{vir}}$ and $M_{\mathrm{vir}}$ is not well-known and was studied in various simulations of structure formation. Eq. \ref{eq3} is used following the halo concentration fit of \citep{Dolag} which is in good agreement with most of N-body simulations proposed in the literature (see \citep{Bullock} and \citep{Eke} as examples). In Eq. \ref{eq3}, $z$ denotes the redshift and $h$ the present day normalized Hubble expansion rate. $c_\mathrm{0}$ and $\alpha$ are the parameters of the halo concentration fit and depend on the cosmological scenario ($c_\mathrm{0}$ = 9.6 and $\mathrm{\alpha = -0.1}$ in a $\Lambda$CDM cosmology).\\
The CMa dSph galaxy is located close to the Galactic disk and suffers from strong tidal disruptions. A reasonably good estimator of its total dark halo mass is then the mass enclosed inside its tidal radius rather than its virial mass. The tidal radius of the CMa dwarf galaxy is calculated via the Roche criterion:
\begin{equation}
\frac{M_{\mathrm{dSph}}(r_\mathrm{t})}{r_\mathrm{t}^{\mathrm{3}}} = \frac{M_{\mathrm{MW}}(d-r_\mathrm{t})}{(d-r_\mathrm{t})^{\mathrm{3}}},\label{tidal}
\end{equation}
where d is the distance of CMa to the center of the MW. $M_{\mathrm{MW}}$(r) denotes the mass of the MW galaxy enclosed in a sphere of radius r. A NFW profile for the MW halo is considered with a concentration parameter equal to 10 and a virial mass of 10$^{12}$ M$_{\odot}$. The total mass of the dSph galaxy is computed by iterative tidal stripping, inserting first $M_{\mathrm{vir}}$ in Eq. \ref{tidal} and computing successively the total halo mass (using Eq. \ref{eq1}) and the tidal radius (using Eq. \ref{tidal}), until the convergence of the procedure is reached. The question is now whether or not tidal forces significantly remodel the internal structure of tidally affected dSph. Discrepant results have been reported in the literature regarding this question \citep{Reed, Stoehr}. Here, it is assumed that tidal forces do not affect the inner part of the density profile so that the initial halo structural parameters are kept constant during the stripping procedure. The remaining mass is typically found to be an order of magnitude lower than the virial mass.\\
The astrophysical term $f^{\mathrm{AP}}$ can then be computed as a function of the halo mass by performing the line-of-sight integration of the CMa dSph squared mass density, according to Eq. \ref{Eq4}. Table \ref{table1} shows the obtained NFW structural parameters for a sample of three dark matter halos with different virial masses. The integral of the squared mass density increases with the halo mass and the $f^{\mathrm{AP}}$ value for a dwarf galaxy of mass 10$^{8}$ M$_{\odot}$ located at an heliocentric distance of 8 kpc is found to be in the right order of magnitude ($f^{\mathrm{AP}}\sim$10$^{25}$ GeV$^{2}$ cm$^{-5}$) compared to \citep{EvansFerrer}.
\placetable{table1}
\subsection{Halo independent constraints on the annihilation signal}
\begin{figure}
\plottwo{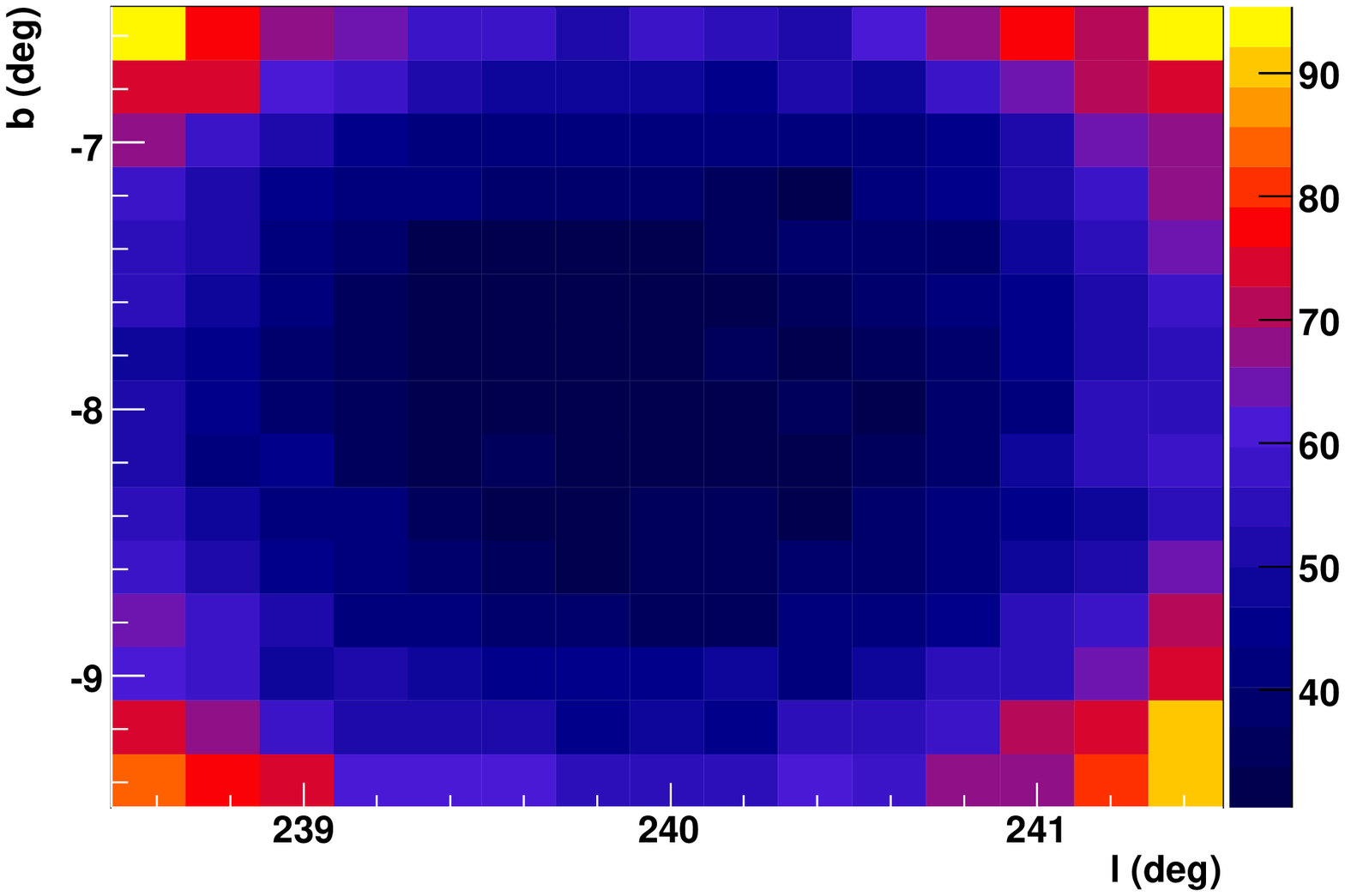}{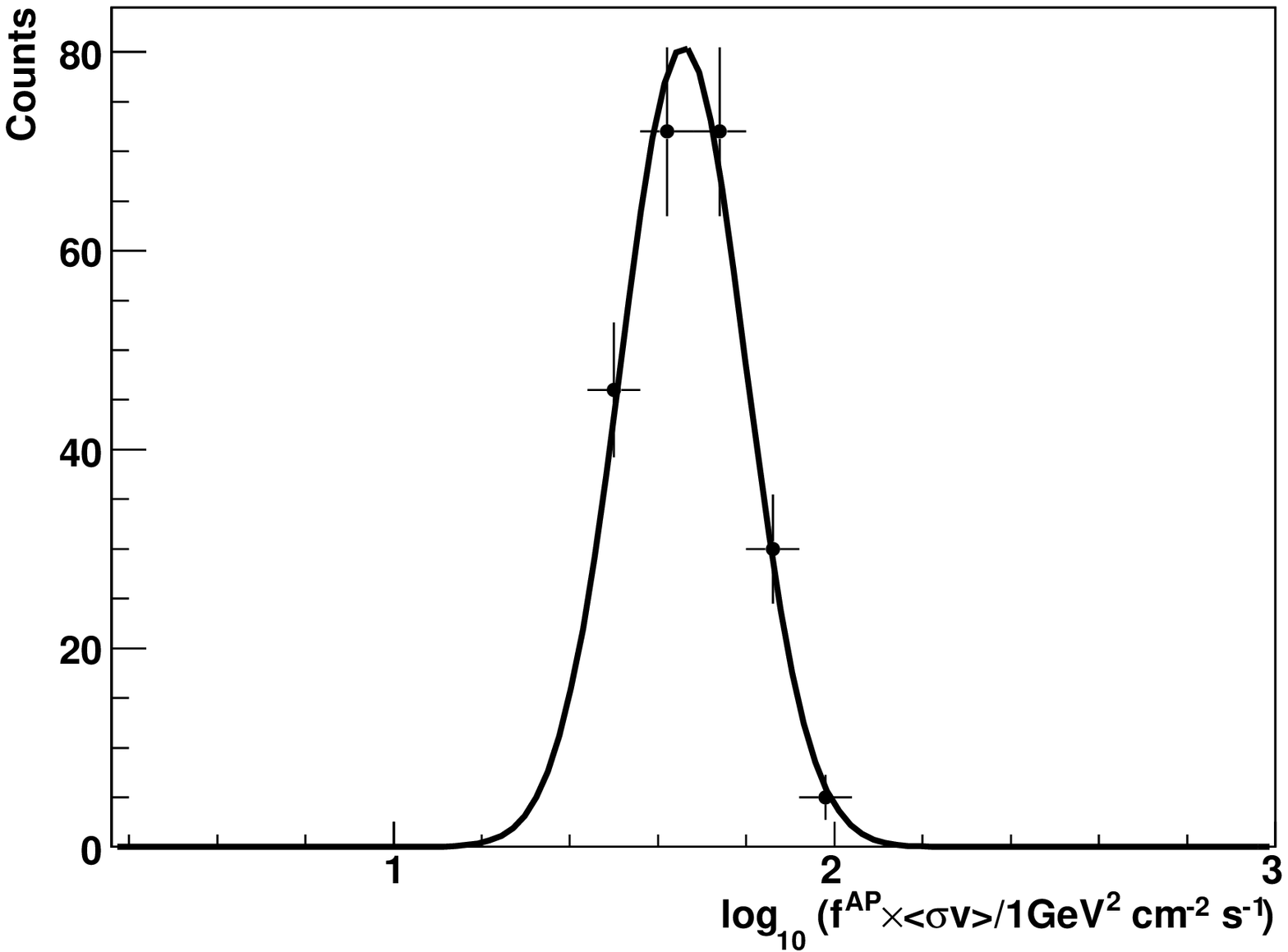}
\caption{(left) 95$\%$ C.L. upper limit map on the value of $\langle\sigma v\rangle$ $\times$ f$^{\mathrm{AP}}$ derived for a 1 TeV neutralino (see text). (right) Distribution of the 95$\%$ C.L. $\langle\sigma v\rangle$ $\times$ f$^{\mathrm{AP}}$ upper limits logarithm derived with the upper limit map from the left panel. The solid line shows the Gaussian fit. The mean value is 1.66 $\pm$ 0.01 and the corresponding variance is 0.14 $\pm$ 0.01.}\label{fig2}
\end{figure}

Using the upper limit $\gamma$ map on the number of $\gamma$-events derived with the $\gamma$ candidate map and the normalized background map (see Sec 3.2 for details), upper limits on the value of $\langle\sigma v\rangle$ $\times$ $f^{\mathrm{AP}}$ can be derived (Eq. \ref{Eq1}, Eq. \ref{Eq2}, Eq. \ref{Eq3}) for various neutralino masses. As an example, the left panel of Fig. \ref{fig2} shows a sky map of the 95$\%$ C.L upper limit values on $\langle\sigma v\rangle$ $\times$ $f^{\mathrm{AP}}$, computed within a pMSSM scenario for a 1 TeV higgsino-like neutralino annihilating in pairs of W and Z gauge bosons. The sensitivity is better toward the center of the FoV, because the H.E.S.S. acceptance decreases on the edges of the camera. H.E.S.S. is sensitive to $\langle\sigma v\rangle$ $\times$ $f^{\mathrm{AP}}$ values of the order of 10$^{2}$ GeV$^{2}$ cm$^{-2}$ s$^{-1}$. Since the core of the CMa dSph can be anywhere in the FoV, one obtains a distribution of the $\langle\sigma v\rangle$ $\times$ $f^{\mathrm{AP}}$ upper limits over the FoV. As shown by the right panel of Fig. \ref{fig2}, this distribution is well fitted by a Gaussian curve. The mean value can be taken as the $\langle\sigma v\rangle$ $\times$ $f^{\mathrm{AP}}$ 95$\%$ C.L upper limit reference value and the corresponding 1$\sigma$ error bar reflects the uncertainties associated to it.\\
Using this procedure, an exclusion curve in the plane ($\langle\sigma v\rangle$ $\times$ $f^{\mathrm{AP}}$, $m_{\mathrm{DM}}$) can be derived. The corresponding curve is shown on the left panel of Fig. \ref{fig3}. The grey shaded areas represent the 1$\sigma$ uncertainties associated with the 1$\sigma$ error bars on the $\langle\sigma v\rangle$ $\times$ $f^{\mathrm{AP}}$ 95$\%$ upper limit.
\placefigure{fig2}
\subsection{Sensitivity to the CMa mass}
\begin{figure}
\plottwo{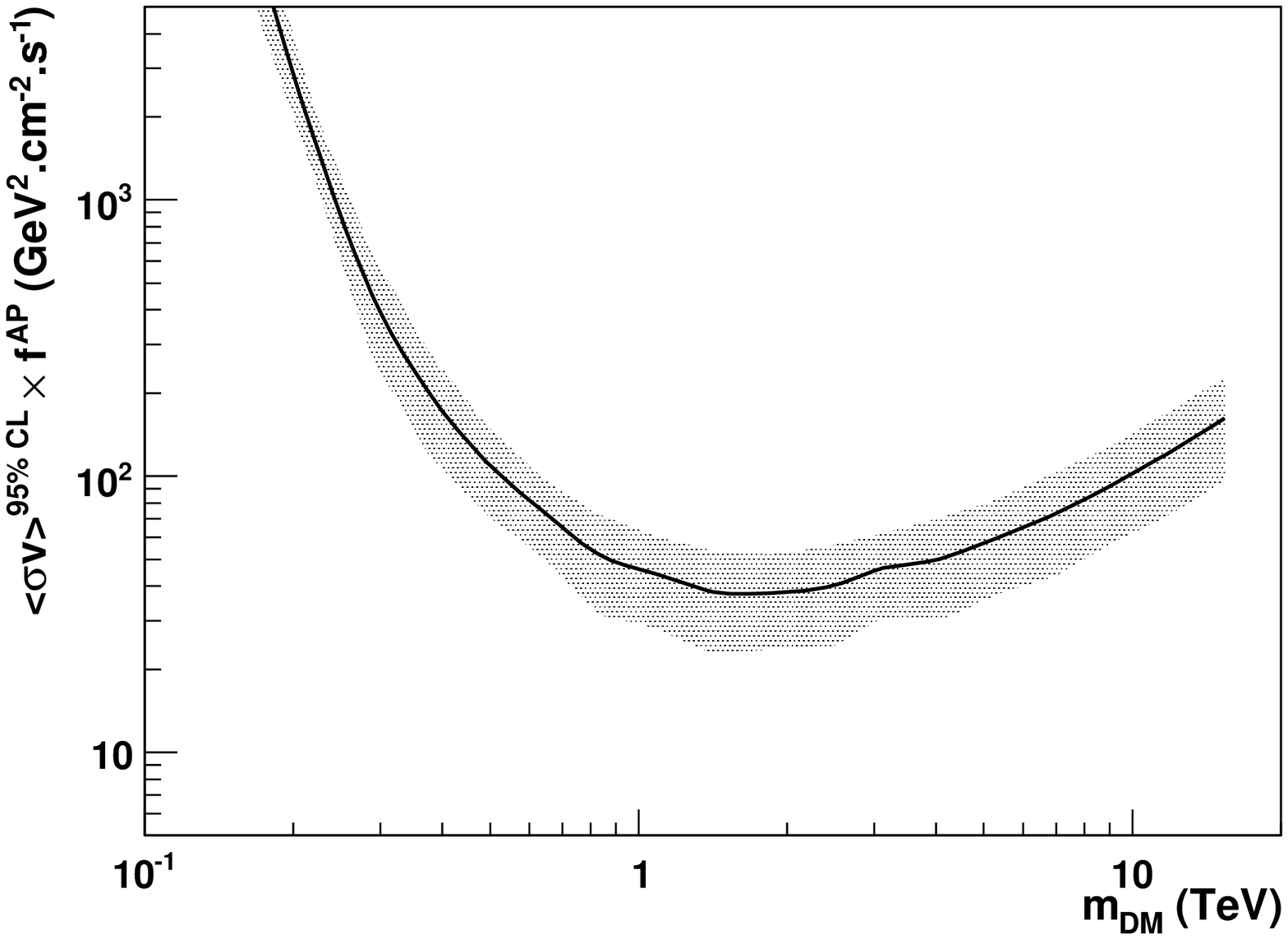}{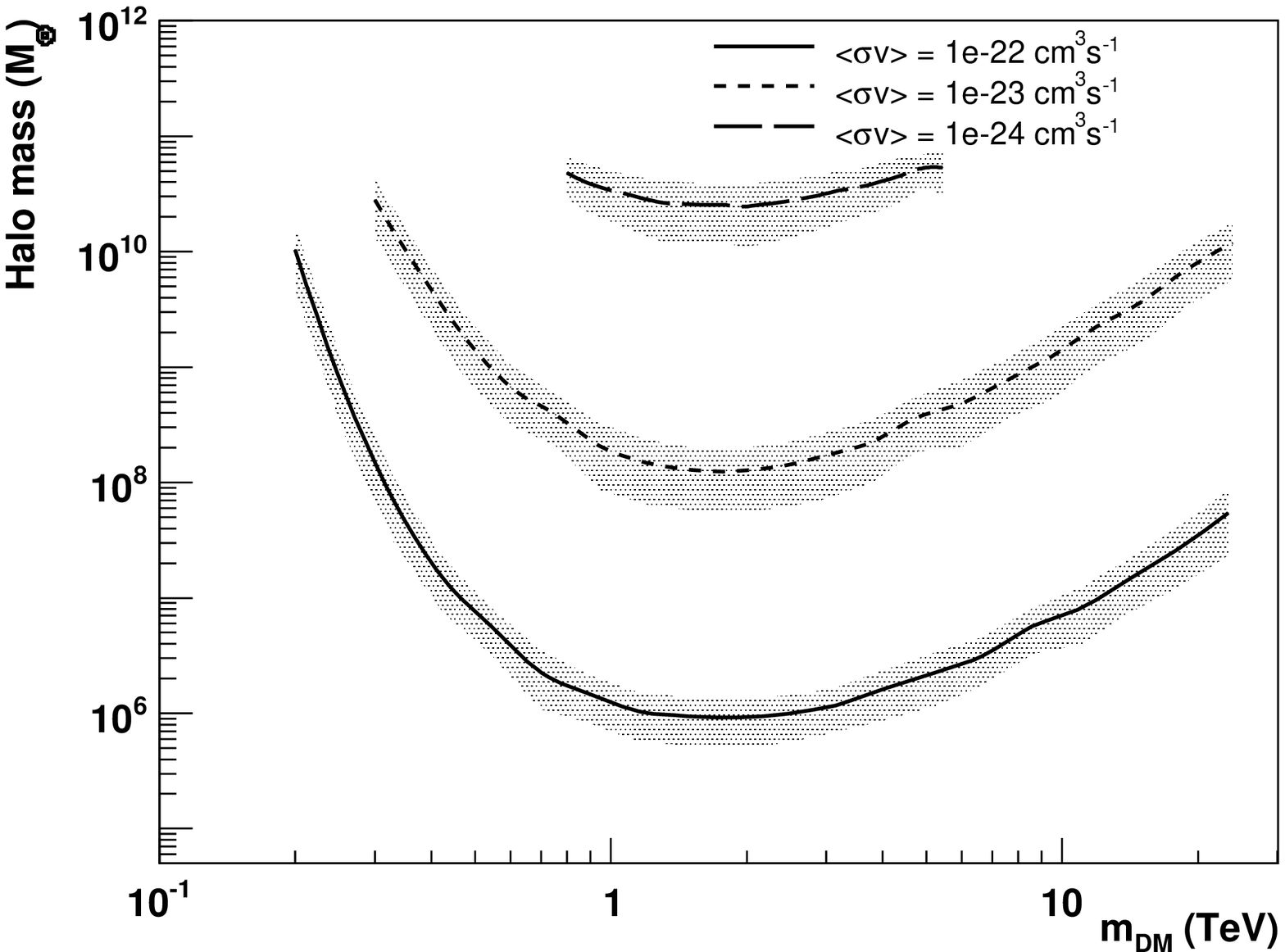}
\caption{(left) Upper limits at 95\% CL on the value of $\langle\sigma v\rangle$ $\times$ f$^{\mathrm{AP}}$ as a function of the DM particle mass in the framework of pMSSM scenarios. (right) Upper limits at 95$\%$ CL on the CMa total mass versus the DM particle mass for different annihilation cross-sections in pMSSM scenarios. The shaded area represents the error bars issued from the 1$\sigma$ error on the $\langle\sigma v\rangle$ $\times$ f$^{\mathrm{AP}}$ distribution Gaussian fits (see text for details).}\label{fig3}
\end{figure}

As the $f^{\mathrm{AP}}$ factor is related to the CMa halo mass (see Sec 4.1), an upper limit on the total CMa mass can be obtained using the previously derived exclusion curve on the $\langle\sigma v\rangle$ $\times$ $f^{\mathrm{AP}}$ quantity and assuming a fixed value for the annihilation cross-section. WIMPs velocity-weighted annihilation cross-sections are expected to be of the order of weak-scale interaction cross-sections. Exclusion curves in the plane ($M_{\mathrm{CMa}}^{\mathrm{95\% C.L}}$,$m_{\mathrm{DM}}$) are then plotted for different annihilation cross-sections within pMSSM scenarios, using the parametrization of \citep{Bergstrom3} for the $\gamma$-ray annihilation spectrum. The corresponding curves are shown on the right panel of Fig. \ref{fig3}. The grey shaded areas represent the 1$\sigma$ uncertainties associated with the 1$\sigma$ error bars on the $\langle\sigma v\rangle$ $\times$ $f^{\mathrm{AP}}$ 95$\%$ upper limit, as described in the previous section. Annihilation cross-sections larger than 10$^{-24}$ cm$^{3}$ s$^{-1}$ are considered here. Lower cross-sections would have been too small  to constrain the CMa dSph mass, e.g. the excluded masses would have been of the order of a MW-sized galaxy for typical velocity-weighted annihilation cross-section of 3$\times$ 10$^{-26}$ cm$^{3}$ s$^{-1}$.
\placefigure{fig3}
\subsection{Sensitivity to the annihilation cross-section of WIMP candidates}
\begin{figure}
\plottwo{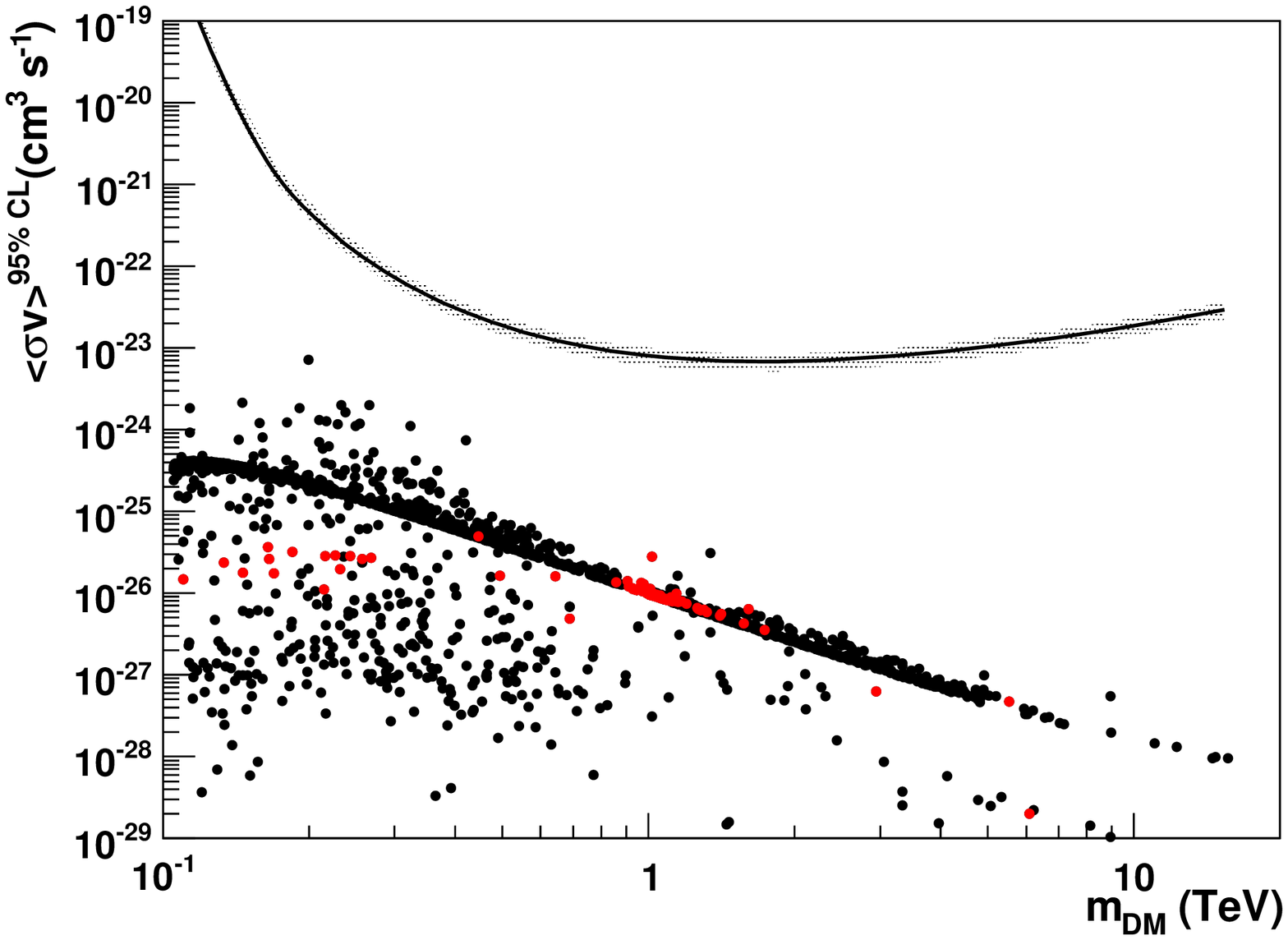}{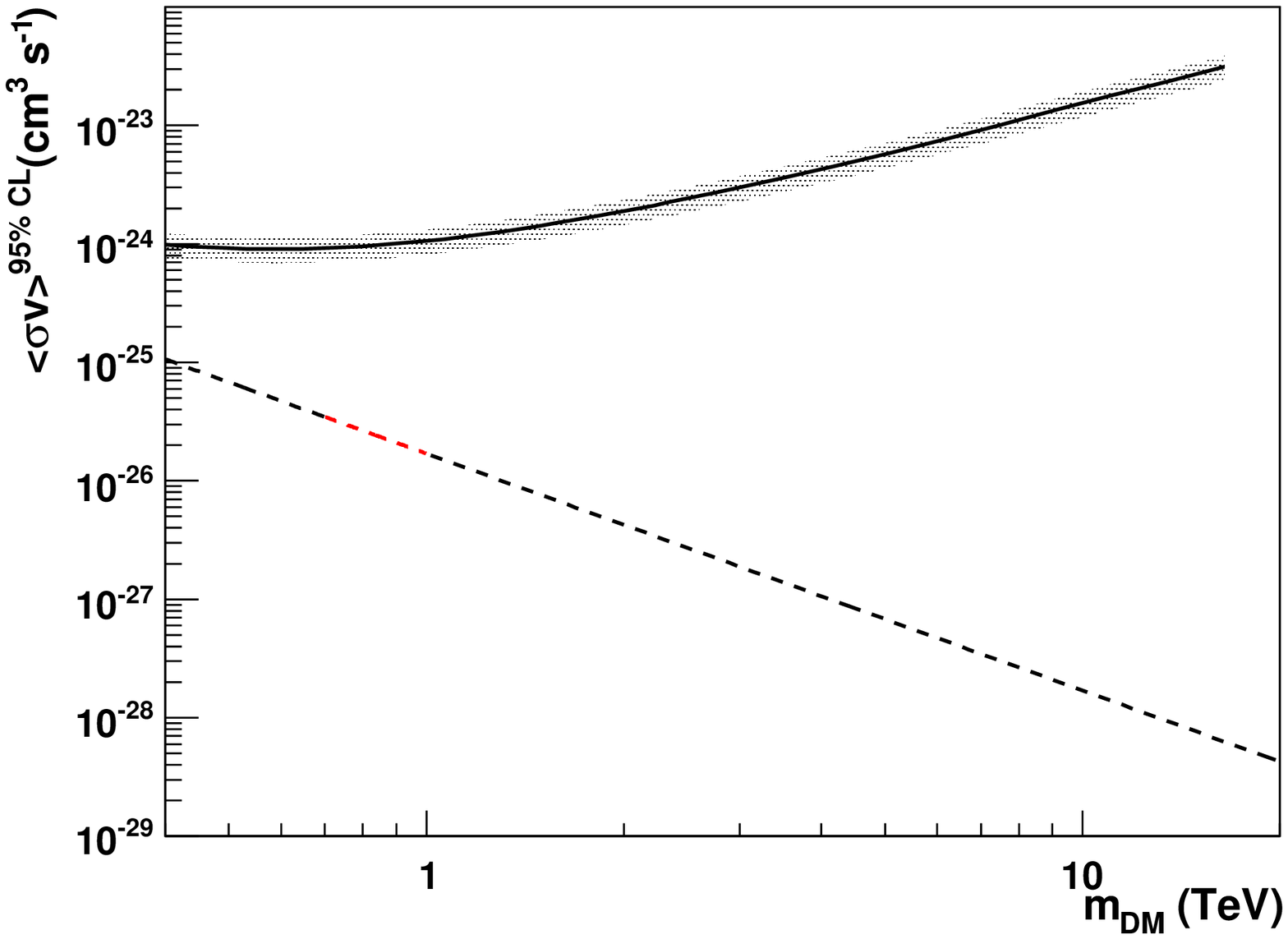}
\caption{Upper limits at 95$\%$ CL on the velocity weighted cross-section as a function the DM particle mass in the case of pMSSM (left panel) and KK (right panel) scenarios, for an assumed CMa total mass of 3 $\times$ 10$^{8}$ M$_{\odot}$. The shaded area represents the 1$\sigma$ error bars on $\langle\sigma v\rangle^{\mathrm{95\% C.L}}$ (see text for details). (left) The pMSSM models are represented by black points, and those giving a CDM relic density in agreement with the measured WMAP+SDSS value are illustrated by red points. (right) The KK models are represented by the black dashed line, and those verifying the WMAP+SDSS constraint on $\Omega_{\mathrm{CDM}}h^{\mathrm{2}}$ are labelled in red.}\label{fig4}
\end{figure}

In this part, the CMa total mass is fixed to be 3 $\times$ 10$^{8}$ M$_{\odot}$ and the corresponding value of the astrophysical contribution $f^{\mathrm{AP}}$ in the expected $\gamma$-ray flux is computed following the procedure described in Sec 4.1 Table \ref{table3} compares the value of the CMa dSph and Sgr dSph structural parameters \footnote{The NFW structural parameter for the Sgr dSph galaxy were estimated in \citep{EvansFerrer} using the velocity dispersion measurements of the Draco dSph galaxy whereas those of the CMa dSph are derived using numerical simulation results (see Sec 4.1)} assuming a NFW profile. The contribution of the astrophysical term f$^{\mathrm{AP}}$ is larger for the CMa dSph than for the Sgr dSph because CMa is closer to the sun. Limits on the velocity-weighted annihilation cross-section $\langle\sigma v\rangle^{\mathrm{95\% C.L}}$ can then be derived as a function of the DM particle mass. Limits are computed in the framework of SUSY and KK models.\\
\placetable{table3}
The SUSY parameters are computed with the micrOMEGAs v1.37 software package \citep{micrOMEGAs2}. Phenomenological MSSM scenarios have been considered. They are characterized by 7 independent parameters: the higgsino mass parameter $\mu$, the common sfermions scalar mass $m_\mathrm{0}$, the Higgs fields vaccum expectation value ratio $tan\beta$, the gaugino mass $M_\mathrm{2}$, the trilinear couplings $A_\mathrm{t}$ and $A_\mathrm{b}$ and the CP-odd Higgs mass $M_\mathrm{A}$. Table \ref{table2} summarizes the region of the pMMSM parameter space scanned to generate the models. The left panel of Fig. \ref{fig4} shows the H.E.S.S. exclusion limits on the velocity weighted cross-section. The black points illustrate the computed pMSSM scenarios and the red points represent those satisfying the WMAP+SDSS constraints on the CDM relic density $\Omega_{\mathrm{CDM}}h^{\mathrm{2}}$ \citep{WMAPSDSS}. $\Omega_{\mathrm{CDM}}h^{\mathrm{2}}$ is allowed to range between 0.09 and 0.11. The H.E.S.S. observations of the CMa dSph allows to exclude velocity weighted cross-sections of the order of 5 $\times$ 10$^{-24}$ cm$^{3}$ s$^{-1}$, comparable with those derived for the Sgr dSph modelled with a cusped NFW profile. The limits obtained are an order of magnitude larger than the velocity-weighted annihilation cross sections of higgsino-like neutralinos.\\
\placetable{table2}
In the case of KK scenarios, predictions for the velocity-weighted cross-section are computed with the formula given in \citep{Baltz}. The expression of $\langle\sigma v\rangle$ is inversely proportional to the squared mass of the lightest Kaluza-Klein (LKP) particle, namely the B$^{(1)}$ particle. Considered KK models that reproduce the CDM relic measured by WMAP and SDSS require a LKP mass ranging from 0.7 TeV to 1 TeV. The right panel of Fig. \ref{fig4} shows the H.E.S.S. limits obtained within these models. The H.E.S.S. observations do not constrain the KK velocity weighted cross-section.
\placefigure{fig4}
\section{Conclusions}
The CMa overdensity is the subject of many debates over whether it is a dwarf galaxy or the warp and flare of the Galactic outer disk. Considering the first scenario, its relative proximity makes it potentially the best region for searches of a DM annihilation signal. However, the lack of observational data prevents the precise modelling of its density profile. Assuming a NFW profile and a mass content of 3 $\times$ 10$^{8}$ M$_{\odot}$ within its tidal radius, typical of dwarf galaxies, H.E.S.S. is close to exclude a few pMMSM scenarios with higgsino-like neutralinos, but does not reach the necessary sensitivity to test models compatible with the WMPA+SDSS constraint on the CDM relic density. In the case of DM made of B$^{(1)}$ particle from KK models with extra dimensions, no constraints are obtained.

\begin{acknowledgements}
The support of the Namibian authorities and of the University of Namibia
in facilitating the construction and operation of H.E.S.S. is gratefully
acknowledged, as is the support by the German Ministry for Education and
Research (BMBF), the Max Planck Society, the French Ministry for Research,
the CNRS-IN2P3 and the Astroparticle Interdisciplinary Programme of the
CNRS, the U.K. Particle Physics and Astronomy Research Council (PPARC),
the IPNP of the Charles University, the Polish Ministry of Science and 
Higher Education, the South African Department of
Science and Technology and National Research Foundation, and by the
University of Namibia. We appreciate the excellent work of the technical
support staff in Berlin, Durham, Hamburg, Heidelberg, Palaiseau, Paris,
Saclay, and in Namibia in the construction and operation of the
equipment.
\end{acknowledgements}

\clearpage


\begin{table}
\begin{center}
\begin{tabular}{cc}
\hline
Cut name & $\gamma$-event cut value\\ 
\hline
\hline
Combined cut & $\leq$ 0.7\\ 
Image charge min. & $\geq$ 60 photo-electrons\\
Reconstructed shower depth min.(rad. length) & -1\\
Reconstructed shower depth max. (rad. length) &  4\\
Reconstructed nominal distance & $\leq$ 2.5$^{\circ}$\\
Reconstructed event telescope multiplicity & $\geq$ 2\\
\hline
\end{tabular}
\end{center}
\caption{List of cuts used in the analysis. The shower depth is the reconstructed primary interaction depth of the particles. The nominal distance is the angular distance of the image barycenter to the center of the camera.}\label{table0}
\end{table}

\begin{table}[!htp]
\begin{center}
\begin{tabular}{cccccc}
\hline
$M_{\mathrm{vir}}$ & $\rho_\mathrm{0}$ & $r_\mathrm{s}$ & $r_\mathrm{t}$ & M($r\leq r_\mathrm{t}$) & $f^{\mathrm{AP}}$ \\
(M$_{\odot}$) & (10$^{8}$M$_{\odot}$ kpc$^{-3}$) & (kpc) & (kpc) & (M$_{\odot}$) & (10$^{24}$GeV$^{2}$ cm$^{-5}$) \\
\hline
\hline 
10$^{6}$ & 4.7 & 0.04 & 0.28  & 3.9 $\times$ 10$^{5}$ & 0.24 \\
10$^{8}$ & 1.3 & 0.28 & 1.17 & 3.1 $\times$ 10$^{7}$ & 2.2 \\
10$^{10}$ & 0.39 & 2.08 & 4.15 & 1.9 $\times$ 10$^{9}$ & 12 \\
\hline
\end{tabular}
\end{center}
\caption{Structural parameters of NFW dark matter halos and associated $f^{\mathrm{AP}}$ values derived with the procedure described in the text, for three different virial masses.}\label{table1}
\end{table}

\begin{table}
\begin{center}
\begin{tabular}{ccccc}
\hline
& Heliocentric distance & $\rho_\mathrm{0}$ & $r_\mathrm{s}$ & $f^{\mathrm{AP}}$\\ 
& (kpc) & (10$^{8}$M$_{\odot}$ kpc$^{-3}$) & (kpc) & (10$^{24}$GeV$^{2}$ cm$^{-5}$) \\
\hline
\hline
CMa & 8 & 1.1 & 0.55 & 5.9\\ 
Sgr & 24 & 1.4 & 0.62 & 2.2\\
\hline
\end{tabular}
\end{center}
\caption{Comparison table of the NFW structural parameters of the CMa and Sgr dSph, with their associated f$^{\mathrm{AP}}$ values, assuming for both a total mass of 3 $\times$ 10$^{8}$ M$_{\odot}$.}\label{table3}
\end{table}

\begin{table}[!htp]
\begin{center}
\begin{tabular}{c}
\hline
pMSSM parameters\\ 
\hline
\hline
100GeV$\leqslant \mu \leqslant$30TeV \\
100GeV$\leqslant m_\mathrm{0} \leqslant$1TeV \\
1.2$\leqslant tan\beta \leqslant$60 \\
10GeV$\leqslant M_\mathrm{2} \leqslant$50TeV \\
-3TeV$\leqslant A_{\mathrm{t,b}} \leqslant$3TeV \\
50GeV$\leqslant M_\mathrm{A} \leqslant$10TeV\\
\hline
\end{tabular}
\end{center}
\caption{Region of the pMSSM parameter space scanned to generate the models. A set of free parameters in the considered range is associated to a pMSSM model.}\label{table2}
\end{table}

\clearpage


\end{document}